\documentstyle[prd,preprint,tighten,aps,eqsecnum,epsfig,%
  amssymb,newlfont]{revtex}           
\newcommand{\mIm}{\,\mathrm{Im}\,}

\begin{document} 
\bibliographystyle{plain} 
\draft

%
\preprint{
\begin{tabular}{r}
hep-ph/9903402 \\ 
UWThPh--1998--63\\
HEPHY--PUB 705 \\ 
\today
\end{tabular}
}
\title{Electron and Neutron Electric Dipole Moments in the 
  Constrained MSSM}
\author{\mbox{A. Bartl}\footnote{email:
  \texttt{bartl@ap.univie.ac.at}}$^{1}$, 
 \mbox{T. Gajdosik}\footnote{email:
  \texttt{garfield@hephy.oeaw.ac.at}}$^{2}$,
 \mbox{W. Porod}\footnote{email:
  \texttt{porod@ap.univie.ac.at}}$^{1}$, 
 \mbox{P. Stockinger}\footnote{email:
  \texttt{stocki@hephy.oeaw.ac.at}}$^{1}$, 
 \mbox{and H. Stremnitzer}\footnote{email:
  \texttt{strem@ap.univie.ac.at}}$^{1}$}
\address{$^{1}$Institut f\"ur Theoretische Physik, 
  Universit\"at Wien, A-1090, Vienna, Austria \\
  $^{2}$ Institut f\"ur Hochenergiephysik der \"Osterreichischen 
  Akademie der Wissenschaften, A-1050, Vienna, Austria }
\maketitle
\begin{abstract}
We analyze the effects of $CP$-violating phases on the electric 
dipole moment (EDM) of electron and neutron in the constrained 
minimal supersymmetric model. We find that the phases $\varphi_{\mu}$ 
and $\varphi_{A_{0}}$ have to be strongly correlated, in particular 
for small values of the SUSY mass parameters. We calculate the 
neutron EDM in two different models, the Quark--Parton Model 
and the Chiral Quark Model. It turns out that the predictions are 
quite sensitive to the model used. We show parameter regions in 
the $M_{0}$-$M_{1/2}$ plane which are excluded by considering 
simultaneously the experimental bounds of both electron and 
neutron EDM, assuming specific values for the phases $\varphi_{\mu}$ 
and $\varphi_{A_{0}}$.
\end{abstract}


%
\section{Introduction} 
\label{sec:intro}

The electric dipole moments (EDMs) of electron and neutron are 
important observables for testing  our ideas of
$CP$-violation. In the standard model (SM), only one
$CP$-violating phase exists in the Cabbibo-Kobayashi-Maskawa 
matrix. The predictions for the EDMs are extremely small, since 
the first nonzero contributions arise at two-loop level.
They are several orders of magnitude smaller than the experimental 
limits \cite{EDM}. Therefore, the EDMs are well 
suited for testing physics beyond the standard model \cite{Barr}.

In supersymmetric (SUSY) extensions of the standard model, additional 
$CP$-violating phases are possible. Moreover, the first nonzero 
contribution to the EDM already shows up at one-loop level. In
particular, in the minimal supersymmetric standard model (MSSM)
complex parameters can be introduced in the mixing matrices of 
squarks, sleptons, charginos and
neutralinos, therefore  yielding more possible sources of
$CP$-violation. In weak--scale SUSY, the 
masses of the lightest supersymmetric particles are expected 
to be between 100 GeV and 1 TeV. In this case, the EDMs can easily 
be much larger then the experimental limits  and yield 
constraints on the $CP$-violating phases and on the other 
parameters of the MSSM. The conclusion would be 
that either the phases are small or the masses are 
large \cite{{liste1},{Oshimo}}. Other arguments like the electroweak 
origin of the cosmological baryon asymmetry (BAU), however, 
would favor large $CP$-violating phases with relatively small 
masses \cite{Carena}. Therefore, more careful analyses of the 
supersymmetric contributions to the EDM are necessary to 
clarify the situation.

A suitable framework for numerical calculations in SUSY is
the constrained MSSM, also called minimal supergravity--inspired 
model (mSUGRA) \cite{Drees95}. In this model universality of the 
soft SUSY--breaking parameters at the grand unification (GUT) scale 
is assumed. Parameters of the model are the common gaugino mass 
$M_{1/2}$, the common scalar mass $M_{0}$, the common trilinear 
scalar coupling parameter $A_{0}$, and 
$\tan\beta = v_{2}/v_{1}$, with $v_{1,2}$ being the 
vacuum expectation values of the two Higgs fields. The number of 
independent complex phases can be reduced to two. The masses at the 
electroweak scale are determined by using renormalization group 
equations (RGEs). Such an approach to constrain the phases has
recently been used in \cite{Garisto} and \cite{Olive}.

An important aspect in the calculation of the SUSY contributions to 
the EDMs of electron (eEDM) and neutron (nEDM) is the fact that 
strong cancellations between the different contributions can occur. 
This has been particularly emphasized in \cite{Nath}, where 
the nEDM and eEDM in mSUGRA with two complex phases have been
analyzed. Due to this cancellations, the bounds on the phases 
are less restrictive then those found in previous analyses.
Additional constraints on the phases originating from the 
cosmological bounds on the relic density of neutralinos have
been studied in \cite{Falk}.
A different point of view has been presented in
\cite{Kane} where a model with seven independent phases 
at the electroweak scale has been assumed. Also in this analysis
it was found that various cancellations between different 
contributions occur, and that in large regions of the parameter 
space the phases are not necessarily small. 

In our paper we analyze the eEDM and the nEDM simultaneously in 
mSUGRA with complex phases $\varphi_{\mu}$ and $\varphi_{A_{0}}$,
which are the phases of the higgsino mass parameter $\mu$ and the
trilinear scalar coupling parameter $A_{0}$. We use RGEs to 
calculate particle masses, couplings, and phases at the electroweak 
scale from the input parameters at the GUT scale. We confirm the 
importance of cancellations. We find that quite general the 
cancellations occur between the two most important contributions, 
which are the chargino and neutralino contribution in the case of 
the eEDM and the chargino and gluino contribution in the case of 
the nEDM. Furthermore, the cancellations are only possible if the 
phases $\varphi_{\mu}$ and $\varphi_{A_{0}}$ are strongly correlated, 
in particular for small SUSY particle masses. In this case 
$\varphi_{\mu}$ is strongly restricted. For the nEDM, there is 
also the problem of evaluating the hadronic matrix element. We use 
two different approaches, one based on the Quark--Parton Model 
\cite{Ellis}, and a second one based on the Chiral Quark Model 
\cite{Manohar}. We find that the predictions for the nEDM are very 
different for the two models used. We show the regions in the 
$M_{0}$-$M_{1/2}$ plane which are excluded by the experimental 
bounds for both EDMs for specific values of the phases 
$\varphi_{\mu}$ and $\varphi_{A_{0}}$. Finally, we also introduce 
an additional phase $\varphi_{3}$ for the gluino mass parameter 
and study its influence. We find that also $\varphi_{3}$ is 
strongly restricted.

In Sec.~\ref{sec:EDM} we give the expressions for the various 
contributions for eEDM and the quark EDMs, including the 
chromoelectric and purely gluonic dimension-six operator. 
We calculate the nEDM in terms of the quark EDMs in the two different 
models. In Sec.~\ref{sec:rge} we determine the phases and MSSM 
parameters at the electroweak scale using the RGEs. In 
Sec.~\ref{sec:analysis} we give the numerical analysis 
of the EDMs within mSUGRA and a discussion of the results. 
A summary is given in Sec.~\ref{sec:summary}. Explicit forms of 
the mass matrices for sfermions, charginos, and neutralinos, as well 
as the expressions for the RGEs are given in the Appendices. 

%
\section{Contributions to the EDM of electron and neutron} 
\label{sec:EDM}
The EDM of a spin-$\frac{1}{2}$ particle is 
the coefficient $d^{f}$ of the effective operator
\begin{equation} 
{\mathcal L}_{E} = -(i/2) d^{f}
\bar{f} \gamma^{5} \sigma_{\mu\nu} f F^{\mu\nu} 
\enspace . 
\end{equation} 
We calculate the supersymmetric contributions to the EDMs of electron 
and quarks at one--loop level. In the case of the electron EDM 
we include chargino--sneutrino and neutralino--selectron loops. In 
the light quark case we include chargino--squark, 
neutralino--squark, and gluino--squark loops. For 
the chromoelectric dipole moments of quarks we include 
chargino--squark, neutralino--squark, and gluino--squark loops, 
whereas the gluonic dimension--six operator gets contributions from 
loops containing top quark, top squark, and gluino. 

The parts of the SUSY Lagrangian that are necessary to calculate the 
one--loop contributions mentioned above are 
\begin{eqnarray} 
 {\mathcal L}_{\bar{f}\,\tilde{\chi}^{0}_{k}\,\tilde{f}_{m}} 
&=& 
 g \,
 \bar{f} \, 
 ( a_{mk}^{\tilde{f}} P_{R} + b_{mk}^{\tilde{f}} P_{L} ) \, 
 \tilde{\chi}^{0}_{k} \, \tilde{f}_{m}
\label{Lagrangianqch0qt}
\enspace ,
\\ 
 {\mathcal L}_{f^{\prime} \,\tilde{\chi}^{+}_{k}\,\tilde{f}_{m}} 
&=& 
  g \, 
  \bar{f}^{\prime} \, 
  ( l_{mk}^{\tilde{f}} P_{R} + k_{mk}^{\tilde{f}} P_{L} ) \, 
  \tilde{\chi}^{+}_{k} \, \tilde{f}_{m}
\label{Lagrangianqschpqt}
\enspace ,
\\
 {\mathcal L}_{\bar{q}\,\tilde{g}\,\tilde{q}_{m}} 
&=& 
 - ( g_{s} / \sqrt{2} ) \,
 \bar{q} \, \lambda^{a}
\left(
   e^{ \frac{i}{2} \varphi_{3}}
   {\mathcal R}^{\tilde{q} \, *}_{m1} P_{R} 
 - e^{-\frac{i}{2} \varphi_{3}}
   {\mathcal R}^{\tilde{q} \, *}_{m2} P_{L} 
\right) \, 
 \tilde{g}^{a} \, \tilde{q}_{m}
\enspace ,
\label{Lagrangianqgtqt}
\end{eqnarray}
where $g$ and $g_{s}$ are the electroweak and strong coupling 
constants, respectively, $P_{L,R} = ( 1 \mp \gamma^{5} ) / 2$,  
$a=1 \ldots 8$ are the gluino color indices, $\lambda^{a}$ are 
the Gell-Mann matrices, $\varphi_{3}$ is the phase of the 
soft--breaking gluino mass. To simplify the notation the quark and 
squark color indices are suppressed. The scalar  
fields $\tilde{f}_{L}$ and $\tilde{f}_{R}$ are linear 
combinations of the mass eigenstates $\tilde{f}_{1,2}$: 
\begin{equation} 
  {\tilde{f}_{1} \choose \tilde{f}_{2}}
= {\mathcal R}^{\tilde{f}} {\tilde{f}_{L} \choose \tilde{f}_{R}} 
\enspace ,
\end{equation} 
where ${\mathcal R}^{\tilde{f}}$ is the unitary diagonalization 
matrix defined in Eq.~(\ref{s2}). Note that 
${\mathcal R}^{\tilde{f}}$ depends on the phases 
$\varphi_{\mu}$ and $\varphi_{A_{0}}$ via the off diagonal entry of 
the squark mass matrix, see Eq. (\ref{sfphase}) and Table 
\ref{tab:phases}. 
The couplings are defined as (we use the notation of \cite{sabine}): 
\begin{mathletters}
\begin{eqnarray}
l^{\tilde{\nu}}_{mj}
&=&
- \delta_{m1} V_{j1}
\enspace , 
\\ 
l^{\tilde{u}}_{mj}
&=&
- {\mathcal R}^{\tilde{u} \, *}_{m1} V_{j1}
+ Y_{u} {\mathcal R}^{\tilde{u} \, *}_{m2} V_{j2}
\enspace , 
\\ 
l^{\tilde{e},\tilde{d}}_{mj}
&=&
- {\mathcal R}^{\tilde{e},\tilde{d} \, *}_{m1} U_{j1}
+ Y_{e,d} {\mathcal R}^{\tilde{e},\tilde{d} \, *}_{m2} U_{j2}
\enspace , 
\end{eqnarray}
\end{mathletters}%
\begin{mathletters}
\begin{eqnarray} 
k^{\tilde{\nu}}_{mj}
&=&
Y_{e} \delta_{m1} U_{j2}^{*}
\enspace , 
\\
k^{\tilde{e}}_{mj}
&=&
0 
\enspace , 
\\ 
k^{\tilde{u}}_{mj}
&=&
Y_{d} {\mathcal R}^{\tilde{u} \, *}_{m1} U_{j2}^{*}
\enspace , 
\\
k^{\tilde{d}}_{mj}
&=&
Y_{u} {\mathcal R}^{\tilde{d} \, *}_{m1} V_{j2}^{*} 
\enspace , 
\end{eqnarray}
\end{mathletters}%
\begin{mathletters}
\begin{eqnarray} 
a^{\tilde{f}}_{mj}
&=&
  {\mathcal R}^{\tilde{f} \, *}_{m1} f^{\tilde{f}}_{Lj}
+ {\mathcal R}^{\tilde{f} \, *}_{m2} h^{\tilde{f}}_{Rj}
\enspace , 
\\ 
b^{\tilde{f}}_{mj}
&=&
  {\mathcal R}^{\tilde{f} \, *}_{m1} h^{\tilde{f}}_{Lj}
+ {\mathcal R}^{\tilde{f} \, *}_{m2} f^{\tilde{f}}_{Rj} 
\enspace , 
\end{eqnarray} 
\end{mathletters}%
\begin{mathletters}
\begin{eqnarray} 
h^{u}_{Lj}
&=&
Y_{u} ( \sin\beta N_{3j} - \cos\beta N_{4j} )
\enspace , 
\\ 
h^{u}_{Rj}
&=&
Y_{u} ( \sin\beta N^{*}_{3j} - \cos\beta N^{*}_{4j} ) = h^{u*}_{Lj}
\enspace , 
\\    
h^{e,d}_{Lj}
&=&
- Y_{e,d} ( \cos\beta N_{3j} + \sin\beta N_{4j} )
\enspace , 
\label{couplings:hsd}
\\
h^{e,d}_{Rj}
&=&
- Y_{e,d} ( \cos\beta N^{*}_{3j} + \sin\beta N^{*}_{4j} ) 
= h^{e,d \, *}_{Lj}
\enspace , 
\end{eqnarray}
\end{mathletters}%
\begin{mathletters}
\begin{eqnarray} 
f^{f}_{Lj}
&=& 
- \left[
    Q_{f} \sin 2\theta_{W} N^{*}_{1j}
  + ( 1 - 2 Q_{f} \sin^{2}\theta_{W} ) N^{*}_{2j} 
  \right] 
  / ( \sqrt{2} \cos\theta_{W} )
\enspace , 
\\ 
f^{f}_{Rj}
&=&
  \left[
    Q_{f} \sin 2\theta_{W} N_{1j}
  + ( - 2 Q_{f} \sin^{2}\theta_{W} ) N_{2j} 
  \right] 
  / ( \sqrt{2} \cos\theta_{W} ) 
\enspace , 
\end{eqnarray}
\end{mathletters}%
\begin{mathletters}
\begin{eqnarray}
Y_{u} 
&=& \frac{m_{u}}{\sqrt{2} \, m_{W} \sin\beta} 
\enspace , 
\\ 
Y_{e,d} 
&=& \frac{m_{e,d}}{\sqrt{2} \, m_{W} \cos\beta} 
\enspace .
\label{couplings:Ys}
\end{eqnarray}
\end{mathletters}%
$Q_{f}$ and $Y_{f}$ are electric and Yukawa couplings of the 
fermion $f$, $\theta_{W}$ is the Weinberg angle, and 
$\tan\beta = v_{2} / v_{1}$ 
is the ratio of the Higgs vacuum expectation values $v_{1}$ and 
$v_{2}$. $U$ and $V$ are the unitary matrices which diagonalize the 
chargino mass matrix, Eq.~(\ref{charmass}).
$N_{\alpha j}$ is the unitary matrix which diagonalizes the complex 
symmetric neutralino mass matrix, Eq.~(\ref{neutmass}).
For diagonalizing we use the singular value decomposition. 

A generic form for the one--loop EDM of spin-$1/2$ particles due to 
exchange of fermions and scalar particles has been worked out 
in~\cite{Grimus}. Extensions of the EDMs to the full electric 
and weak dipole moment form factors for the top quark have been 
given in \cite{Bartl}.
A non--vaninshing EDM demands a change in chirality
of the external fermion and involves the imaginary parts of the 
couplings. In the following we give the complete analytic expressions 
for the individual one--loop contributions. We have compared our 
results with \cite{{Nath},{Kane}} and found agreement.

\subsection{Chargino Contribution} 
\label{sec:EDM:chargino}

The chargino contribution to the EDM of the fermion $f$ is given by 
\begin{equation} 
\frac{1}{e}
d^{f}_{\tilde{\chi}^{+}}
=
\frac{\alpha}{4\pi\sin^{2}\theta_{W}} 
\sum_{m,j=1}^{2} \mIm [\Gamma^{f}_{mj} ] 
  \frac{m_{\tilde{\chi}^{+}_{j}}}{m^{2}_{\tilde{f}^{\prime}_{m}}} 
  \left( 
    Q_{f^{\prime}} 
    B( \frac{m^{2}_{\tilde{\chi}^{+}_{j}}}
            {m^{2}_{\tilde{f}^{\prime}_{m}}} )
  + ( Q_{f} - Q_{f^{\prime}} ) 
    A( \frac{m^{2}_{\tilde{\chi}^{+}_{j}}}
            {m^{2}_{\tilde{f}^{\prime}_{m}}} )
  \right)
\enspace , 
\label{charginocontribution}
\end{equation} 
where $\alpha = e^{2}/(4\pi)$ and $e = g \sin\theta_{W}$. 
$f^{\prime}$ is the isospin partner of $f$ in the SU(2)--doublet. 
Neglecting the mass of the external fermions (in our case electron, 
up, down, and strange quark) the functions A and B have the 
simple form~\cite{Grimus}
\begin{eqnarray} 
A(r) &=& \frac{1}{2 ( 1 - r )^{2}} 
  \left( 3 - r + \frac{2 \ln r}{1 - r} \right)
\enspace ,
\label{afunction}
\\ 
B(r) &=& \frac{1}{2 ( 1 - r )^{2}} 
  \left( 1 + r + \frac{2 r \ln r}{1 - r} \right)
\enspace .  
\label{bfunction}
\end{eqnarray} 
The first and second term in Eq.~(\ref{charginocontribution}) are due 
to the Feynman diagrams Fig.~1a and Fig.~1b, respectively. The 
expressions $\mIm[ \Gamma^{f}_{mj} ]$ are given by:
\begin{eqnarray}
\label{elecGamma} 
\mIm[ \Gamma^{e}_{mj} ]
&=&
\mIm[ Y_{e} U_{j2} V_{j1} ] \delta_{m1}
\enspace ,
\end{eqnarray} 
\begin{eqnarray}
\label{upGamma} 
\mIm[ \Gamma^{u}_{mj} ]
&=&
\mIm[ Y_{u} V_{j2} {\mathcal R}^{\tilde{d}}_{m1}
  ( U_{j1} {\mathcal R}^{\tilde{d} \, *}_{m1} 
  - Y_{d} U_{j2} {\mathcal R}^{\tilde{d} \, *}_{m2} ) ]
\\ 
&=&
  (1/2) Y_{u} 
\left( 
  ( 1 - (-1)^{m} \cos 2\theta_{\tilde{d}} ) 
  \mIm[ U_{j1} V_{j2} ] 
+ Y_{d} (-1)^{m} \sin2\theta_{\tilde{d}}
  \mIm[ U_{j2} V_{j2} e^{i\varphi_{\tilde{d}}} ] 
\right) 
\enspace ,
\nonumber\end{eqnarray} 
and 
\begin{eqnarray}
\label{downGamma} 
\mIm[ \Gamma^{d}_{mj} ]
&=&
\mIm[ Y_{d} U_{j2} {\mathcal R}^{\tilde{u}}_{m1}
  ( V_{j1} {\mathcal R}^{\tilde{u} \, *}_{m1} 
  - Y_{u} V_{j2} {\mathcal R}^{\tilde{u} \, *}_{m2} ) ]
\\ 
&=&
  (1/2) Y_{d} 
\left( 
  ( 1 - (-1)^{m} \cos 2\theta_{\tilde{u}} ) 
  \mIm[ U_{j2} V_{j1} ] 
+ Y_{u} (-1)^{m} \sin2\theta_{\tilde{u}}
  \mIm[ U_{j2} V_{j2} e^{i\varphi_{\tilde{u}}} ] 
\right) 
\enspace .
\nonumber\end{eqnarray}

\subsection{Neutralino Contribution} 
\label{sec:EDM:neutralino}

The neutralino contribution to the fermion EDM is given by 
\begin{equation} 
\frac{1}{e}
d^{f}_{\tilde{\chi}^{0}}
=
- \frac{Q_{f}}{8\pi}
  \frac{\alpha}{\sin^{2}\theta_{W}}
\sum_{k=1}^{4} \sum_{m=1}^{2} \eta^{f}_{mk} 
  \frac{m_{\tilde{\chi}^{0}_{k}}}{m^{2}_{\tilde{f}_{m}}}
  B( \frac{m^{2}_{\tilde{\chi}^{0}_{k}}}{m^{2}_{\tilde{f}_{m}}} ) 
\end{equation} 
where 
\begin{eqnarray}
\label{eta} 
\eta^{f}_{mk} 
&=&
(-1)^{m} \sin2\theta_{\tilde{f}}
  \mIm[ ( (h^{f}_{Lk})^{2} - f^{f}_{Lk} f^{f*}_{Rk} ) 
        e^{-i \varphi_{\tilde{f}}}] 
\nonumber\\ && 
- ( 1 - (-1)^{m} \cos2\theta_{\tilde{f}})
  \mIm[ h^{f}_{Lk} f^{f*}_{Lk} ]
\nonumber\\ && 
- ( 1 + (-1)^{m} \cos2\theta_{\tilde{f}})
  \mIm[ h^{f}_{Lk} f^{f}_{Rk} ] 
\enspace .
\end{eqnarray} 

\subsection{Gluino Contribution} 
\label{sec:EDM:gluino}

The gluino contribution to the quark EDM is given by 
\begin{eqnarray} 
\frac{1}{e}
d^{q}_{\tilde{g}}
&=&
- \frac{2 \alpha_{s}}{3\pi} \sum^{2}_{k=1} 
  \mIm [ e^{i \varphi_{3}} 
         {\mathcal R}^{\tilde{q}}_{k2} 
         {\mathcal R}^{\tilde{q} \, *}_{k1} ]
  \frac{m_{\tilde{g}}}{m^{2}_{\tilde{q}_{k}}} 
  Q_{q}\enspace
  B( \frac{m^{2}_{\tilde{g}}}{m^{2}_{\tilde{q}_{k}}} )
\nonumber\\ 
&=&
  \frac{\alpha_{s}}{3\pi}
  \sin ( \varphi_{3} - \varphi_{\tilde{q}} ) 
  \sin 2\theta_{\tilde{q}} 
  \sum^{2}_{k=1} 
  (-1)^{k} \frac{m_{\tilde{g}}}{m^{2}_{\tilde{q}_{k}}} 
  Q_{q}\enspace
  B( \frac{m^{2}_{\tilde{g}}}{m^{2}_{\tilde{q}_{k}}} )
\enspace , 
\label{gluinocontribution}
\end{eqnarray} 
where $\alpha_{s}=g_{s}^{2}/4\pi$ and $m_{\tilde{g}}$ is the gluino 
mass. 

\subsection{Quark Chromoelectric Dipole Moment and %
Gluonic Dimension--Six Operator} 
\label{sec:EDM:chromo}
 
The quark chromoelectric dipole moment is defined as the 
coefficient $\hat{d}^{q}$ in the effective operator 
\begin{equation} 
{\mathcal L}_{C}
=
- (i/2) \hat{d}^{q} 
  \bar{q} \sigma_{\mu\nu} \gamma^{5} (\lambda^{a}/2) q \, G^{a\mu\nu} 
\enspace .
\end{equation} 
The chromoelectric dipole moment has also chargino, neutralino, and 
gluino contributions. They are given by~\cite{Nath} 
\begin{equation}
\hat{d}^{q}_{\tilde{\chi}^{+}}
=
- \frac{g_{s}\alpha}{4 \pi \sin^{2}\theta_{W}}
\sum^{2}_{m,j=1} 
  \mIm [ \Gamma^{q}_{mj} ]
  \frac{m_{\tilde{\chi}^{+}_{j}}}{m^{2}_{\tilde{q}_{m}}} 
  B( \frac{m_{\tilde{\chi}^{+}_{j}}}{m^{2}_{\tilde{q}_{m}}} ) 
\enspace , 
\end{equation}
\begin{equation}
\hat{d}^{q}_{\tilde{\chi}^{0}}
=
- \frac{g_{s}\alpha}{8 \pi \sin^{2}\theta_{W}}
\sum_{m=1}^{2} \sum_{k=1}^{4} 
  \eta^{q}_{mk} 
  \frac{m_{\tilde{\chi}^{0}_{k}}}{m^{2}_{\tilde{q}_{m}}} 
  B( \frac{m_{\tilde{\chi}^{0}_{k}}}{m^{2}_{\tilde{q}_{m}}} ) 
\enspace , 
\end{equation}
and 
\begin{eqnarray} 
\hat{d}^{q}_{\tilde{g}}
&=&
- \frac{g_{s}\alpha_{s}}{4\pi}
\sum^{2}_{k=1} 
  \mIm [ e^{i \varphi_{3}} 
         {\mathcal R}^{\tilde{q}}_{k2} 
         {\mathcal R}^{\tilde{q} \, *}_{k1} ]
  \frac{m_{\tilde{g}}}{m^{2}_{\tilde{q}_{k}}} 
  C( \frac{m^{2}_{\tilde{g}}}{m^{2}_{\tilde{q}_{k}}} ) 
\nonumber\\ 
&=&
  \frac{g_{s}\alpha_{s}}{8\pi} 
  \sin ( \varphi_{3} - \varphi_{\tilde{q}} ) 
  \sin 2\theta_{\tilde{q}} 
\sum^{2}_{k=1} 
  (-1)^{k} \frac{m_{\tilde{g}}}{m^{2}_{\tilde{q}_{k}}} 
  C( \frac{m^{2}_{\tilde{g}}}{m^{2}_{\tilde{q}_{k}}} )
\enspace , 
\label{chromo:gluino}
\end{eqnarray}
where
\begin{equation}
\label{Cfunc}
C(r) = 3 A(r) - (1/3) B(r) 
\enspace .
\end{equation} 
The Wilson coefficient $d_{G}$ of the $CP$--violating gluonic 
dimension--six operator is defined through
\begin{equation}
{\cal{L}}_{G} = - (1/6) d_{G} 
G_{\mu\nu a} G_{b}^{\nu\rho} \tilde{G}_{\rho c}^{\mu} f_{abc}
\enspace . 
\end{equation}
The leading nontrivial contribution to $d_{G}$ in the MSSM is 
given by a two--loop diagram involving top, scalar top, and 
gluino \cite{{Nath},{Dai}}: 
\begin{equation}
d_{G} =
 \frac{3 \alpha_{s}^{2} g_{s} m_{t} }{32 \pi^{2}} 
 \sin\varphi_{\tilde{t}} \sin 2\theta_{\tilde{t}} 
 \frac{m_{\tilde{t}_{1}}^{2} - m_{\tilde{t}_{2}}^{2}}
      {m^{5}_{\tilde{g}}} 
 \enspace
 H \left( \frac{m_{\tilde{t}_{1}}^{2}}{m^{2}_{\tilde{g}}}, 
          \frac{m_{\tilde{t}_{2}}^{2}}{m^{2}_{\tilde{g}}},  
          \frac{m_{t}^{2}}{m^{2}_{\tilde{g}}}
   \right) \enspace .
\end{equation} 
The definition of the two--loop function $H$ can be found 
in \cite{Dai}. 

\subsection{EDM of electron and neutron} 
\label{sec:EDM:el-and-neu}

Having defined the contributions from the individual Feynman diagrams, 
we can now write down the total EDM of the electron as the sum of 
neutralino and chargino contributions: 
\begin{equation}
d^{e} = d^{e}_{\tilde{\chi}^{+}} + d^{e}_{\tilde{\chi}^{0}} 
\enspace .
\end{equation}

In order to obtain the EDM of the neutron in terms of the 
quark EDMs, a specific description of the neutron as quark 
bound state is needed. Throughout this paper we 
use two different approaches. 

1.) The relativistic Quark--Parton Model: 
In this model, the contributions of the quarks to the nEDM are 
given in terms of quantities $\Delta_{q}$ \cite{Ellis}, 
which are measured in polarized lepton--nucleon scattering:
\begin{equation}
  d^{n}
= \eta^{E} 
\left( \Delta_{u} d^{d} 
     + \Delta_{d} d^{u} 
     + \Delta_{s} d^{s} 
\right) 
\enspace ,
\label{quark-parton-edm}
\end{equation} 
where the individual quark contributions are again given in terms of 
chargino, neutralino, and gluino contributions
\begin{equation}
  d^{q}
= d^{q}_{\tilde{\chi}^{+}} 
+ d^{q}_{\tilde{\chi}^{0}} 
+ d^{q}_{\tilde{g}}
\enspace .
\end{equation}
As already stated, the $\Delta_{q}$ are the measured contributions of 
the quark $q$ to the spin of the proton; to use them for the neutron 
we have taken advantage of a simple isospin relation. For definiteness 
we use the 
values given in Ref.~\cite{EMC}: \mbox{$\Delta_{u} = 0.746$}, 
\mbox{$\Delta_{d} = -0.508$}, and \mbox{$\Delta_{s} = -0.226$}.
The QCD correction factor $\eta^{E}$ takes into account that the quark 
EDM analysis is done at the electroweak scale and hence has to be 
evolved down to the hadronic scale with the help of RGEs. 
We use $\eta^{E}=1.53$ as given in Ref.~\cite{Arnowitt}.

2.) The Chiral Quark Model: 
This model is based on the effective chiral quark theory given in 
Ref. \cite{Manohar}. The contribution of the quark EDMs to the nEDM is 
given by the nonrelativistic $SU(6)$ coefficients 
\begin{equation}
d^{n} = (4/3) d^{d} - (1/3) d^{u} 
\enspace .
\end{equation}
The quark EDMs in this model are given by contributions of all quark 
and gluon operators (to leading order in $\alpha_{s}$) with the 
proper dimensional rescaling. This yields 
\begin{equation}
d^{q} 
= \eta^{E} 
  ( d^{q}_{\tilde{\chi}^{+}} + d^{q}_{\tilde{\chi}^{0}}
  + d^{q}_{\tilde{g}} ) 
+ \eta^{C} \frac{e}{4\pi} 
  ( \hat{d}^{q}_{\tilde{\chi}^{+}}
  + \hat{d}^{q}_{\tilde{\chi}^{0}}
  + \hat{d}^{q}_{\tilde{g}} )
+ \eta^{G} \frac{e\Lambda_{SB}}{4\pi} d_G
\enspace .
\label{chiral-quark-edm}
\end{equation}
$\eta^{E}$, $\eta^{C}$, and $\eta^{G}$ are the QCD correction factors 
due to RGEs, whereas $\Lambda_{SB}$ is the scale of chiral 
symmetry breaking in QCD; we use $\eta^{E}=1.53$~\cite{Arnowitt}, 
$\eta^{C} \simeq \eta^{G} \simeq 3.4$ (as used in \cite{Nath}), 
and $\Lambda_{SB} \simeq 1.19$~GeV \cite{Manohar}.

%
\section{Determination of the MSSM parameters and phases}
\label{sec:rge}

The formulas for the EDMs, when evaluated in the MSSM with complex 
parameters in its most general form, contain too many free parameters. 
In order to study the constraints 
of the EDMs on the phases and mass parameters we have to reduce the 
number of free parameters by further theoretical assumptions. 
Therefore, we assume universality conditions for gaugino, sfermion, 
and Higgs mass parameters and the trilinear couplings 
\begin{eqnarray}
M_{0} 
&:=& 
  M_{\tilde E_{i}} 
= M_{\tilde L_{i}} 
= M_{\tilde D_{i}} 
= M_{\tilde Q_{i}} 
= M_{\tilde U_{i}} 
= m_{H_{1}} 
= m_{H_{2}} 
\label{eq:boundary1}
\enspace , 
\\
M_{1/2} 
&:=& 
  M_{1} = M_{2} = M_{3} 
\enspace ,
\label{eq:gaugino-boundary}
\\
A_{0} 
&:=& 
  A_{e_{i}} = A_{d_{i}} = A_{u_{i}}
\label{eq:boundary2}
\end{eqnarray}
at the GUT scale $M_{GUT}$ \cite{Drees95}, where $i=1,2,3$ is the 
generation index. We determine the parameters at the electroweak 
scale with the help of the RGEs as given in \cite{castano94}. 

At the electroweak scale the following parameters can be complex: the 
trilinear couplings $A_{f_{i}}$, the gaugino mass parameters $M_{k}$, 
and the Higgs parameters $\mu$ and $B$. The product $\mu B$ and the 
gaugino mass parameter $M_{2}$ can be made real by redefinition of 
the fields. $|\mu|$ and $B$ are determined by requiring the correct 
electroweak symmetry breaking: 
\begin{eqnarray}
  |\mu|^{2} 
&=& 
  \frac{ ( m_{H_{1}}^{2} + \Delta T_{1} ) 
       - ( m_{H_{2}}^{2} + \Delta T_{2} ) \tan^{2}\beta }
       { \tan^{2}\beta - 1 }
- \frac{1}{2} m_{Z}^{2} 
\enspace , 
\label{min1a} 
\\
  2 \mu B 
&=& 
  ( m_{H_{1}}^{2} + m_{H_{2}}^{2} + 2 |\mu|^{2} 
    + \Delta T_{1} + \Delta T_{2} ) \sin 2\beta 
\enspace ,
\label{min2a} 
\end{eqnarray}
where $\Delta T_{1,2}$ denote the leading one--loop corrections to 
the tadpole equations stemming from top, scalar top, bottom, and 
sbottom contributions \cite{{castano94},{barger94},{Helmut}}. The 
phase of $\mu$, $\varphi_{\mu}$, remains a free parameter. 
$\varphi_{\mu}$ can be specified at any scale, because it does not 
evolve with the corresponding RGE up to two loops \cite{martin94}. 
In order to determine the phases at the electroweak scale we assume 
$M_{1/2}$ real, and $A_{0}$ and $\mu$ complex at the GUT scale. Note 
that at one--loop level only the phase difference between the phases 
of $A_{0}$ and $M_{1/2}$ is physically relevant. We summarize the 
complex phases entering the mass matrices in Table~\ref{tab:phases}. 

We use the following procedure for determining the soft SUSY--breaking 
parameters at the electroweak scale. We specify the gauge
couplings, $\tan \beta$, and the Yukawa couplings of the third 
generation at the electroweak scale. We take $A_{0}$, $M_{0}$, 
$M_{1/2}$ at $M_{GUT}$ with 
Eqs.~(\ref{eq:boundary1})--(\ref{eq:boundary2}) as boundary 
conditions. The RGEs are given in the $\overline{\mbox{DR}}$ 
scheme. We evolve the RGEs for the gauge couplings at 
two--loop level from $Q = m_{Z}$ to $Q = M_{GUT}$ which is
determined by the condition $g_{1} = g_{2}$. We evolve the RGEs for 
the Yukawa couplings at the one--loop level, because they enter the 
RGEs of the gauge couplings at two--loop level. We take into account 
threshold effects by including step functions for the coefficients of 
the beta functions (see e.g. \cite{castano94}). For simplicity we 
assume that there is no mixing between the generations. We then evolve 
the RGEs for the soft SUSY--breaking parameters from $M_{GUT}$ to 
$m_{Z}$. The mass parameters $M_{j}$ are decoupled from the RGEs if 
$M_{j}(Q) = Q$ is satisfied. We calculate $|\mu|$ and $B$ by requiring 
correct electroweak symmetry breaking Eqs.~(\ref{min1a}) and 
(\ref{min2a}). The corrections are sensitive to the relative phases 
between the $A$ parameters and $\mu$. This phase dependence may change 
$|\mu|$ by a few GeV, which is in the range of the error expected by 
neglecting the other contributions to the one--loop corrected tadpoles 
\cite{{barger94},{Helmut}}. We iterate the complete procedure until 
the parameters vary less than 1\%. 

For the discussion in the next Section it is convenient to have the 
following approximations for the parameters at the electroweak scale 
at hand (the exact formulas for the one--loop results are given in 
Appendix~\ref{sec:AppRGE}). With $\alpha_{GUT} = 1/24$ and 
$M_{GUT} = 2.38 \times 10^{16}$~GeV we get: 
\begin{mathletters}
\begin{eqnarray}
\label{semp}
M^{2}_{\tilde L} &\simeq& M_{0}^{2} + 0.52 M^{2}_{1/2} 
\enspace , 
\\
M^{2}_{\tilde E} &\simeq& M_{0}^{2} + 0.15 M^{2}_{1/2} 
\enspace , 
\\
M^{2}_{\tilde Q} &\simeq& M_{0}^{2} + 6.7 M^{2}_{1/2} 
\enspace , 
\\
\label{sqmp}
M^{2}_{\tilde U} &\simeq& M^{2}_{\tilde D} 
                \simeq M_{0}^{2} + 6.2 M^{2}_{1/2} 
\enspace , 
\end{eqnarray}
\end{mathletters}%
\begin{mathletters}
\begin{eqnarray}
\label{seMi}
M_{1} = M^{\prime} &\simeq& 0.41 M_{1/2} 
\enspace , 
\\
M_{2} = M &\simeq& 0.82 M_{1/2} 
\enspace , 
\\
M_{3} &\simeq& 2.82 M_{1/2} 
\enspace , 
\end{eqnarray}
\end{mathletters}%
\begin{mathletters}
\begin{eqnarray}
\label{Atparameter}
A_{t} &\simeq& (1-y) \, A_{0} - 2 \, M_{1/2} 
\enspace , 
\\
\label{Auparameter}
A_{u} &\simeq& 
\left(1 - \textstyle \frac{y}{2} \right) \, A_{0} - 2.8 \, M_{1/2} 
\enspace , 
\\
\label{Adparameter}
A_{d} &\simeq& A_{0} - 3.6 \, M_{1/2} 
\enspace , 
\\
\label{Aeparameter}
A_{e} &\simeq& A_{0} - 0.7 \, M_{1/2}
\enspace , 
\end{eqnarray}
\end{mathletters}%
where y varies between 0.85 and 1 for $40 > \tan\beta > 1$. 
Eqs.~(\ref{semp})--(\ref{sqmp}) are only valid for the first and 
second generation. 
Note that Eqs.~(\ref{Atparameter})--(\ref{Aeparameter}) have strong
implications for the $A$ parameters at the electroweak scale. If one 
takes, for example, $M_{1/2}$ real and $A_{0}$ imaginary, 
$A_{0} = i A$, at $M_{GUT}$ then one obtains the values given in
Table~\ref{tab:weak-phases}.

%
\section{EDM Analysis within mSUGRA}
\label{sec:analysis}

In this Section we investigate the EDM of electron and neutron in the 
framework of mSUGRA with complex parameters. As outlined in 
Sec.~\ref{sec:rge}, this model is completely specified by six 
parameters: $M_{0}$, $M_{1/2}$, $|A_{0}|$, $\tan\beta$ and the phases 
$\varphi_{A_{0}}$ and $\varphi_{\mu}$. The experimental bounds 
obtained in \cite{EDM} are 
$|d^{e}| \le d^{e}_{\mathrm{exp}} = 4.3 \times 10^{-27} \, e$cm and 
$|d^{n}| \le d^{n}_{\mathrm{exp}} = 1.1 \times 10^{-25} \, e$cm.

For the eEDM we have two supersymmetric contributions stemming from 
neutralino and chargino exchange, Figs.~1a and 1b, respectively. The 
chargino contribution depends explicitly on the phase $\varphi_{\mu}$, 
the dependence on $\varphi_{A_{0}}$ comes only through the RGEs 
and is very weak. The neutralino contribution depends explicitly 
on $\varphi_{\mu}$ and $\varphi_{A_{0}}$. 
In the major part of the parameter space the 
chargino contribution dominates. The reasons are: 
(i) The loop function $A(r)$, Eq.~(\ref{afunction}), 
entering in the chargino contribution, is larger than $B(r)$, 
Eq.~(\ref{bfunction}), which enters the neutralino contribution.  
(ii) The neutralino contribution is proportional to the selectron 
mixing angle $\sin 2\theta_{\tilde{e}}$, which is usually rather 
small. 

In Fig.~2 we show $d^{e}_{\tilde{\chi}^{0}}$, the neutralino 
contribution of the eEDM, as a function of the $CP$--violating 
phases $\varphi_{\mu}$ and $\varphi_{A_{0}}$ with the other 
parameters fixed: $M_{0} = 150$~GeV, $M_{1/2} = 200$~GeV, 
$|A_{0}| = 450$~GeV, and $\tan\beta = 3$. As can be seen, 
the neutralino contribution alone already exceeds the experimental 
limit. The calculated eEDM is below the experimental limit
only if cancellations between chargino and neutralino 
contributions occur. In this case the eEDM 
depends significantly on the phase $\varphi_{A_{0}}$ if either 
$|\varphi_{\mu}| \ll |\varphi_{A_{0}}|$ or 
$|A_{e}| \gtrsim |\mu|\tan\beta$. In the first case the chargino 
contribution is small because it is proportional to 
$\sin \varphi_{\mu}$, therefore, the neutralino contribution can be 
of the same order of magnitude as the chargino contribution.
In the second case the relevant phase in the neutralino contribution 
is determined by the off--diagonal element of the selectron mixing 
matrix Eq.~(\ref{sfphase}). In the mSUGRA model the absolute value of 
$\mu$ is fixed by the condition of radiative electroweak symmetry 
breaking Eq.~(\ref{min1a}). It turns out that $|\mu|$ has always 
roughly the same order of magnitude as $|A_{e}|$ in the parameter 
region considered. Note that the neutralino contribution depends not 
only on the phase of $( A_{e} - \mu^{*} \tan\beta )$, 
Eq.~(\ref{sfphase}), but also directly on $\varphi_{\mu}$ via 
the neutralino mixing matrix, as can be seen in Eqs.~(\ref{eta}), 
(\ref{couplings:hsd})--(\ref{couplings:Ys}).

Due to the cancellation mechanism between chargino and neutralino 
contribution it is not straightforward to conclude which mSUGRA 
parameter values and phases are excluded by the experimental 
upper bound of the eEDM. To answer this question we show 
in Fig.~3 the regions in the $M_{0}$-$M_{1/2}$ plane that are 
allowed by the experimental limit on the eEDM for different 
values of the phase $\varphi_{\mu}$. In doing so we have taken 
$\varphi_{A_{0}} = \pi/2$, which is the maximal phase 
difference between $M_{1/2}$ and $A_{0}$ at the GUT scale, 
$\tan\beta = 3$, and $|A_{0}| = 3 M_{0}$. For example, choosing 
$\varphi_{\mu} = -0.1$, the region in the $M_{0}$-$M_{1/2}$ plane 
to the left of the dashed--dotted line is excluded. As can be seen, 
the parameters $M_{0} = 120$~GeV and $M_{1/2} = 160$~GeV are allowed 
and give relatively light SUSY particle masses (for illustration: 
$m_{\tilde{\chi}^{0}_{1}} = 58$~GeV, 
$m_{\tilde{\chi}^{\pm}_{1}} = 106$~GeV, 
$m_{\tilde{\nu}_{e}} = 157$~GeV, 
$m_{\tilde{e}_{1}} = 160$~GeV, 
$m_{\tilde{e}_{2}} = 163$~GeV).
Taking $\varphi_{\mu} = -0.54$, only values of $(M_{1/2},M_{0})$ to 
the right of the solid line are allowed which, for example, means 
$M_{1/2} \gtrsim 0.9$~TeV if $M_{0} = 1.5$~TeV or 
$M_{1/2} \gtrsim 1.5$~TeV if $M_{0} = 0.7$~TeV. In this case rather 
heavy SUSY particles are predicted, (\mbox{i. e.} 
$m_{\tilde{\chi}^{0}_{1}} > 398$~GeV, 
$m_{\tilde{\chi}^{\pm}_{1}} > 764$~GeV, 
$m_{\tilde{\nu}_{e}} > 1170$~GeV, 
$m_{\tilde{e}_{1}} > 1073$~GeV, 
$m_{\tilde{e}_{2}} > 1171$~GeV).
The bending in the dotted line for $\varphi_{\mu} = -0.18$ is caused 
by the cancellation mechanism between chargino and neutralino 
contributions. The grey area is excluded, because the condition of 
radiative electroweak symmetry breaking is not fulfilled. 

Up to now we have only considered the eEDM. Now we
consider the eEDM and the nEDM simultanously. Taking into account 
also the experimental upper limit on the nEDM will enlarge the 
excluded parameter region. The predicted value for the nEDM depends 
strongly on the neutron model which relates the nEDM to the EDM of 
its constituents. To demonstrate this fact we calculate the nEDM in 
the Quark--Parton Model and in the Chiral Quark Model as described in 
Sec.~\ref{sec:EDM:el-and-neu}. Also for the nEDM to fulfill the 
experimental bounds it is necessary that strong cancellations between 
the different contributions occur. 

Another way to show the systematics of these cancellations is to plot 
the allowed region in the $\varphi_{\mu}$-$\varphi_{A_{0}}$ plane. 
In the Figs.~4, 5, and 8 we consider rather small mSUGRA parameters: 
\mbox{$M_{0} = 150$~GeV}, \mbox{$M_{1/2} = 200$~GeV}, 
\mbox{$|A_{0}| = 450$~GeV}, and \mbox{$\tan\beta = 3$}. 
In all $\varphi_{\mu}$-$\varphi_{A_{0}}$ plots (Figs.~4, 7, and 8),
the allowed values of the phases are within the small bands between 
the lines. In Figs.~4, 6, 7, and 8 we discuss the eEDM together 
with the nEDM. As can be seen from the dotted lines for the allowed 
region of the eEDM in Figs.~4a and 4b, $\varphi_{\mu}$ is bounded,
$|\varphi_{\mu}| \lesssim 0.1$, whereas $\varphi_{A_{0}}$ is 
essentially unrestricted. However, the two phases have to be strongly 
correlated: for every $\varphi_{A_{0}}$, $\varphi_{\mu}$ can only 
vary in an interval $\Delta\varphi_{\mu} \lesssim 0.01$. Taking into 
account only the chargino contribution, one would obtain the 
restriction $|\varphi_{\mu}| \lesssim 0.01$. 

In Fig.~4a we show the experimentally allowed regions for the 
eEDM and the nEDM, calculated in the Quark--Parton Model. For 
the paramters chosen and the measured spin densities of the 
proton~\cite{EMC} \mbox{$\Delta_{u} = 0.746$}, 
\mbox{$\Delta_{d} = -0.508$}, and \mbox{$\Delta_{s} = -0.226$}, the 
allowed band in the $\varphi_{\mu}$-$\varphi_{A_{0}}$ plane of the 
nEDM lies within the allowed band of the eEDM. In this case the nEDM 
is more restrictive. For the values of the spin densities taken, the 
nEDM and the eEDM have opposite signs. In Fig.~4b we plot the allowed 
band in the $\varphi_{\mu}$-$\varphi_{A_{0}}$ plane of the nEDM, 
calculated in the Chiral Quark Model, and compare it to the eEDM. 
They have the same sign. As one can see, in this 
case only a very small region of the parameter space is not excluded 
by experiment: $|\varphi_{\mu}| \lesssim 0.01$ and 
$|\varphi_{A_{0}}| \lesssim 0.15$. (All phases have to be 
understood modulo $\pi$.) 

In Figs.~5a and 5b we demonstrate the cancellation effects that play 
an essential role in the calculation of the nEDM. We choose the 
relation $\varphi_{\mu} = - (\pi/30) \cdot \sin\varphi_{A_{0}}$, which 
guarantees that the nEDM calculated in the Quark--Parton Model 
fulfills the experimental bound. We show the different contributions 
to the nEDM for the same parameters as in Fig.~4a. 
In Fig.~5a we show the corresponding chargino, neutralino, and gluino 
contributions. As can be seen, there is a strong cancellation between 
chargino and gluino contributions: each of the two contributions 
is approximately 18 times bigger than the whole nEDM. In Fig.~5b we 
show the up, down, and strange quark contributions to the nEDM. 
Again, cancellations between the individual quark contributions 
occur. It turns out, that the strange quark contribution is the most 
important one, as noted in \cite{Ellis}. Therefore, it may turn out 
that an accurate measurement of the nEDM can also become a test of 
the spin structure of the neutron in the Quark--Parton Model. 

In the Chiral Quark Model the cancellations occur for up and down 
quark seperately. There are large cancellations between 
$d^{n}_{\tilde{\chi}^{+}}$ and $d^{n}_{\tilde{g}}$, between 
$\hat{d}^{n}_{\tilde{\chi}^{+}}$ and $\hat{d}^{n}_{\tilde{g}}$, and
also between the resulting sums of this cancellations 
(see Eq.~(\ref{chiral-quark-edm})). The purely gluonic dimension--six 
operator does not exceed the experimental limit 
by itself, however, it can further reduce the total nEDM. 
For the eEDM the cancellation between chargino and neutralino 
contribution exhibits the same behaviour as shown in Fig.~5a, where 
the neutralino contribution in the eEDM plays the same role as the 
gluino contribution in the nEDM. 

In Figs.~6a and 6b we show the regions in the $M_{0}$-$M_{1/2}$ plane 
which are excluded by simultanous consideration of the experimental 
limits on eEDM and nEDM. Fig.~6a is for the Quark--Parton Model and 
Fig.~6b is for the Chiral Quark Model. In both plots we choose the 
following values for the phases: 
\mbox{$\varphi_{A_{0}} = -\pi/10$} (dashed lines), 
\mbox{$\varphi_{A_{0}} = \pi/5$} (dotted lines),
\mbox{$\varphi_{A_{0}} = \pi/2$} (dashed--dotted lines),
\mbox{$\varphi_{\mu} = -\pi/10$} (thin lines), and 
\mbox{$\varphi_{\mu} = -\pi/30$} (thick lines). 
Only values of $(M_{1/2},M_{0})$ to the right of the corresponding 
lines are allowed by eEDM and nEDM simultanously. 
In Fig.~6a there are only four lines, because for 
\mbox{$\varphi_{\mu} = -\pi/10$}, \mbox{$\varphi_{A_{0}} = -\pi/10$}
and 
\mbox{$\varphi_{\mu} = -\pi/10$}, \mbox{$\varphi_{A_{0}} = \pi/5$} 
the parameter region \mbox{$M_{0}, M_{1/2} \lesssim 1.5$~TeV} is 
excluded by experiment. The Quark--Parton Model is in general more 
restrictive than the Chiral Quark Model. However, in the 
Quark--Parton Model much smaller pairs of mass parameters are allowed, 
for example \mbox{$M_{0} = 150$~GeV} and \mbox{$M_{1/2} = 200$~GeV}. 
As can be seen, the 
strongest cancellation effects are found for $|\varphi_{A_{0}}|=\pi/2$ 
and 
$\mathrm{sign}\,\varphi_{A_{0}} = -\,\mathrm{sign}\,\varphi_{\mu}$. 
This is also observed in \cite{Kane}. If 
$\varphi_{A_{0}}$ and $\varphi_{\mu}$ have the same sign, the 
exclusion is more or less indepentent of $M_{0}$. 

Our numerical investigation of the nEDM 
includes the contributions of the one--loop gluino, chargino and 
neutralino exchange diagrams for the electric dipole operators. 
In the Chiral Quark Model we also include the 
chromoelectric dipole operators and the contribution of the purely 
gluonic dimension--six operator. In the following we want to 
discuss qualitatively which contributions are important to understand 
the behavior of the nEDM and its dependence on the mSUGRA parameters. 

The dominant contributions to the nEDM come from the chargino and 
gluino exchange diagrams of the quark EDMs. It is remarkable that
the chargino contribution is almost independent of the phase 
$\varphi_{A_{0}}$. This is due to the fact that the second terms 
of Eqs.~(\ref{upGamma}) and (\ref{downGamma}) are suppressed by the 
Yukawa couplings $Y_{u,d}$ which are very small for light quarks. 
The gluino contribution, Eq.~(\ref{gluinocontribution}), 
depends on both phases, $\varphi_{\mu}$ and 
$\varphi_{A_{0}}$, since it is proportional to the off--diagonal 
element of the squark mass matrix, 
$m_{q} \left( A_{q} - \mu^{*} \Theta (\beta) \right)$, 
(see Eqs.~(\ref{s1}) and (\ref{sfphase})). 
The neutralino contributions to the quark 
EDMs are very small in contrast to the eEDM.

In the Chiral Quark Model the down quark contribution is the most 
important one, because the EDM is proportional to $(4d_{d} - d_{u})$. 
Moreover, for the chargino contribution we have  
$Y_{d} = (m_{d}/m_{u}) \tan\beta \, Y_{u} \gtrsim 6 Y_{u}$
if $\tan\beta \gtrsim 3$. The gluino contribution to the down quark 
is proportional to $m_{d} \mIm[A_{d} - \mu^{*} \tan\beta]$, whereas 
the up quark EDM contains the factor 
$m_{u} \mIm[A_{u} - \mu^{*} \cot\beta]$. Taking 
into account that $|\mu|$ and $|A_{q}|$ have the  
same order of magnitude, we make the following observations: 
The down quark EDM depends mainly on $\mu$. The up quark EDM 
is dominated by the term proportional to $A_{u}$ and is suppressed by 
a factor $(m_{u}/m_{d}) \cot\beta$ compared to the down quark term. 

The chromoelectric contributions (see Sec.~\ref{sec:EDM:chromo}) 
are suppressed by a factor $g_{s}/(4\pi)$ compared to the 
electric dipole operator and, in general, they are less important. 
In the case where $M_{0} > M_{1/2}$ the loop function $C$,  
Eq.~(\ref{Cfunc}), entering $\hat{d}^{q}_{\tilde{g}}$, 
Eq.~(\ref{chromo:gluino}), can compensate this suppression factor 
$g_{s}/(4\pi)$. It also turns out that the contribution of the purely 
gluonic dimension--six operator is very small in the parameter region 
considered. 

In order to see how the restrictions on $\varphi_{\mu}$ and 
$\varphi_{A_{0}}$ depend on the other parameters we also discuss a 
scenario with $|A_{0}| = M_{0}$. In Fig.~7 we show regions in the 
$\varphi_{\mu}$-$\varphi_{A_{0}}$ plane, allowed by the experimental 
bounds on eEDM and nEDM in this case. We calculate the nEDM in the 
Quark--Parton Model with \mbox{$|A_{0}| = M_{0} = 150$~GeV} and the 
other parameters as in Fig.~4a. We find that the phase 
$\varphi_{A_{0}}$ is less important than in the previous scenario 
($|A_{0}| = 3 M_{0}$). The allowed values of $\varphi_{\mu}$ are 
reduced roughly by a factor $1/3$ compared to Fig.~4a, thereby 
suggesting a linear dependence of the allowed values on $|A_{0}|$ 
keeping the other parameters fixed. Furthermore, the value of 
$\tan\beta$ effects the results in a similar way, because it enters 
in the off--diagonal element of the sfermion mixing matrix, 
Eqs.~(\ref{s1})--(\ref{sTheta}). 
This element is only important for the gluino contribution
to the nEDM and the neutralino contribution to the eEDM. As can be
seen in Fig.~7, the bands in the $\varphi_{\mu}$-$\varphi_{A_{0}}$
plane, allowed by the eEDM and the nEDM in the Quark--Parton Model,
overlap similarly as in Fig.~4a.

In order to study the restrictions imposed by the universality 
conditions at the GUT scale, we modify the universality condition for 
the gaugino mass parameters, Eq.~(\ref{eq:gaugino-boundary}). 
We still assume $M_{1/2} := M_{1} = M_{2} = |M_{3}|$, but introduce 
an additional phase $\varphi_{3}$ for the mass parameter $M_{3}$ at 
the GUT scale. We show in Fig.~8 the bands in the 
$\varphi_{\mu}$-$\varphi_{A_{0}}$ plane, allowed by the eEDM and the
nEDM in the Quark--Parton Model, where we take for $\varphi_{3}$ 
the values $0$, $\pi/10$, and $\pi/5$. We take the other parameters 
as in Fig.~4a. The eEDM depends on 
$\varphi_{3}$ only via the RGEs, therefore, this dependence 
is very weak. Comparing the band of the eEDM (dotted 
line) with the bands of the nEDM for values of $\varphi_{3}$ different 
from zero, one can see that $\varphi_{3}$ is strongly 
restricted by experiment. A further possibility would be to 
introduce an additional phase $\varphi_{1}$ for the $U(1)$ gaugino 
mass parameter $M_{1}$. This phase will enter the eEDM and the nEDM. 
It is expected that $\varphi_{1}$ will change the restrictions on 
$\varphi_{3}$ in a similar way as the phase $\varphi_{A_{0}}$ 
changes the restrictions on $\varphi_{\mu}$. 

%
\section{Summary}
\label{sec:summary}

We have studied the eEDM and the nEDM in the framework of mSUGRA with 
complex parameters. We have found that $\varphi_{\mu}$ is strongly 
restricted by the experimental bounds. Moreover, we have found 
that the phases $\varphi_{\mu}$ and $\varphi_{A_{0}}$ have to be 
strongly correlated, in particular for small values of the SUSY 
mass parameters, so that strong cancellations between the different 
contributions occur. For the experimentally allowed values of the 
eEDM, the chargino contribution has to be cancelled by the neutralino 
contribution. The nEDM is dominated by the chargino and gluino 
contributions. The predictions for the nEDM depend 
very sensitively on the model which is used for the neutron.  We have
used the Quark--Parton Model and the Chiral Quark Model to calculate 
the nEDM. We have presented parameter regions in the 
$M_{0}$-$M_{1/2}$ plane which are excluded by simultanous 
consideration of the experimental bounds on the eEDM and the 
nEDM for different values of the phases $\varphi_{\mu}$ and 
$\varphi_{A_{0}}$. 

%
\section*{Acknowledgements} 
This work has been supported by the 'Fonds zur F\"orderung der 
wissenschaftlichen Forschung' of Austria, projects no. P10843--PHY
and  no. P13139--PHY, and by 'Acciones Integradas'. We are grateful 
to E.~Christova, J.~Ellis, W.~Grimus, and W.~Majerotto for very 
useful discussions.


\newpage
\appendix
 
%
\section{Sfermion Mass Matrix} 

The sfermion mass matrices are given by 
\begin{equation}
M^{2}_{\tilde{f}}
=
\left(\begin{array}{cc} 
  M^{2}_{\tilde{f}_{LL}} 
 & e^{-i \varphi_{\tilde{f}}} M^{2}_{\tilde{f}_{LR}} 
\\ 
  e^{i \varphi_{\tilde{f}}} M^{2}_{\tilde{f}_{LR}} 
 & M^{2}_{\tilde{f}_{RR}}
\end{array}\right) 
\enspace ,
\label{s1} 
\end{equation}
where
\begin{eqnarray}
  M^{2}_{\tilde{f}_{LL}} 
&=& 
  M^{2}_{L\tilde{f}} 
+ ( T^{3}_{I} - Q_{f} \sin^{2}\theta_{W} ) \cos 2\beta \, m_{Z}^{2} 
+ m_{f}^{2} 
\enspace ,
\\
  M^{2}_{\tilde{f}_{RR}} 
&=& 
  M^{2}_{R\tilde{f}}
+ Q_{f} \sin^{2}\theta_{W} \cos 2\beta \, m_{Z}^{2}
+ m_{f}^{2} 
\enspace ,
\\
  M^{2}_{\tilde{f}_{LR}} 
&=&
  m_{f} | A_{f} - \mu^{*} \Theta(\beta) | 
\enspace ,
\\
\label{sfphase}
  \varphi_{\tilde{f}}
&=& 
  \arg [ A_{f} - \mu^{*} \Theta(\beta) ]
\enspace ,
\end{eqnarray} 
with 
\begin{equation} 
\Theta(\beta)=\cases{\cot\beta & for $T^{3}_{I} =  \frac{1}{2}$ \cr 
                     \tan\beta & for $T^{3}_{I} = -\frac{1}{2}$ \cr} 
\enspace . 
\label{sTheta}
\end{equation} 
The eigenvalues are given by 
\begin{equation} 
2 m^{2}_{1,2} 
= ( M^{2}_{\tilde{f}_{LL}} + M^{2}_{\tilde{f}_{RR}} ) 
\mp \sqrt{ ( M^{2}_{\tilde{f}_{LL}} - M^{2}_{\tilde{f}_{RR}} )^{2}
         + 4 ( M^{2}_{\tilde{f}_{LR}} )^{2}} 
\enspace ,
\end{equation} 
with $m^{2}_{1} \le m^{2}_{2}$. 
We parametrize the mixing matrix ${\mathcal R}^{\tilde{f}}$ so that  
\begin{equation}
\label{s2} 
\left(\begin{array}{c} 
  \tilde{f}_{1} \\ \tilde{f}_{2} 
\end{array}\right) 
=
{\mathcal R}^{\tilde{f}}
\left(\begin{array}{c} 
  \tilde{f}_{L} \\ \tilde{f}_{R} 
\end{array}\right) 
=
\left(\begin{array}{cc} 
  e^{\frac{i}{2} \varphi_{\tilde{f}}} \cos \theta_{\tilde{f}} 
 & e^{-\frac{i}{2} \varphi_{\tilde{f}}} \sin \theta_{\tilde{f}} 
\\ 
  - e^{\frac{i}{2} \varphi_{\tilde{f}}} \sin \theta_{\tilde{f}} 
 & e^{-\frac{i}{2} \varphi_{\tilde{f}}} \cos \theta_{\tilde{f}} 
\end{array}\right) 
\left(\begin{array}{c} 
  \tilde{f}_{L} \\ \tilde{f}_{R} 
\end{array}\right) 
\enspace ,
\end{equation} 
where $\varphi_{\tilde{f}}$ is given in Eq.~(\ref{sfphase}) and 
\begin{eqnarray} &&
\cos\theta_{\tilde{f}}
=
\frac{-M^{2}_{\tilde{f}_{LR}}}{\Delta} 
\leq 0
\enspace , \quad 
\sin\theta_{\tilde{f}}
=
\frac{M^{2}_{\tilde{f}_{LL}} - m^{2}_{1}}{\Delta}
\geq 0
\enspace , 
\nonumber \\ &&
\Delta^{2} 
=
  ( M^{2}_{\tilde{f}_{LR}} )^{2} 
+ ( m^{2}_{1} - M^{2}_{\tilde{f}_{LL}} )^{2}
\enspace .
\end{eqnarray}


\section{Chargino Mass Matrix}

The chargino mass matrix 
\begin{equation}\label{charmass} 
M^{\tilde{\chi}^{+}}_{\alpha\beta} =
\left(
\begin{array}{cc}
  M                        & m_{W} \sqrt{2} \sin\beta  \\
  m_{W} \sqrt{2} \cos\beta & \mu      
\end{array}
\right)
\end{equation} 
can be diagonalized by the biunitary transformation 
\begin{equation} 
U^{*}_{j\alpha} M^{\tilde{\chi}^{+}}_{\alpha\beta} V^{*}_{k\beta} 
= m_{\tilde{\chi}_{j}^{+}} \delta_{jk}
\enspace ,
\end{equation} 
where $U$ and $V$ are unitary matrices such that  
$m_{\tilde{\chi}_{j}^{+}}$ are positiv and 
$m_{\tilde{\chi}_{1}^{+}} < m_{\tilde{\chi}_{2}^{+}}$. 


\section{Neutralino Mass Matrix}

We define $N_{\alpha j}$ as the unitary matrix which makes the complex 
symmetric neutralino mass matrix diagonal with positiv diagonal 
elements: 
\begin{equation} 
N_{\alpha j} M^{\tilde{\chi}^{0}}_{\alpha\beta} N_{\beta k}
= m_{\tilde{\chi}^{0}_{j}}\delta_{jk}  
\enspace ,
\end{equation} 
where $m_{\tilde{\chi}^{0}_{j}} < m_{\tilde{\chi}^{0}_{k}}$ for $j<k$. 
In the basis \cite{bfmo89}: 
\begin{equation} 
\psi_{\alpha} = 
\{ - i\tilde{\gamma},-i\tilde{Z},\tilde{H}^{a},\tilde{H}^{b} \} 
\enspace ,
\end{equation} 
the complex symmetric neutralino mass matrix has the form 
\begin{equation}\label{neutmass} 
M^{\tilde{\chi}^{0}}_{\alpha\beta} =
\left(
\begin{array}{cccc}
m_{\tilde{\gamma}} &        m_{az} &      0 &   0 \\ 
            m_{az} & m_{\tilde{z}} &  m_{Z} &   0 \\
0 & m_{Z} &   \mu \sin 2\beta & - \mu \cos 2\beta \\ 
0 &     0 & - \mu \cos 2\beta & - \mu \sin 2\beta
\end{array}
\right)
\enspace ,
\end{equation} 
where 
\begin{eqnarray}
m_{\tilde{\gamma}} 
&=& M \sin^{2}\theta_{W}
 + M^{\prime} \cos^{2}\theta_{W}
\enspace ,
\nonumber \\
m_{\tilde{z}} 
&=& M \cos^{2}\theta_{W}
 + M^{\prime} \sin^{2}\theta_{W}
\enspace ,
\\
m_{az} 
&=& \sin\theta_{W} \cos\theta_{W} ( M - M^{\prime} ) 
\enspace .
\nonumber
\end{eqnarray}

%
\section{Solutions of the one--loop RGEs}
\label{sec:AppRGE}

The solutions of the one--loop RGEs (as given in \cite{castano94}) 
for the soft SUSY--breaking parameters are given by 
\begin{eqnarray}
  M_{i}(t) 
&=& 
  \frac{M_{1/2}}{1 + \beta_{i} t} 
\\
  M^{2}_{\tilde{E}_{1}}(t) 
&=& 
  M^{2}_{0}
+ \frac{\alpha_{GUT} M^{2}_{1/2}}{4 \pi} \frac{6}{5} f_{1}(t) 
\\
  M^{2}_{\tilde{L}_{1}}(t) 
&=& 
  M^{2}_{0} + \frac{\alpha_{GUT} M^{2}_{1/2}}{4 \pi} 
  \left( \frac{3}{2} f_{2}(t) + \frac{3}{10} f_{1}(t) \right) 
\\
  M^{2}_{\tilde{D}_{1}}(t) 
&=& 
  M^{2}_{0} + \frac{\alpha_{GUT} M^{2}_{1/2} }{4 \pi}
  \left( \frac{8}{3} f_{3}(t) + \frac{2}{15} f_{1}(t) \right) 
\\
  M^{2}_{\tilde{U}_{1}}(t) 
&=& 
  M^{2}_{0} + \frac{\alpha_{GUT} M^{2}_{1/2}}{4 \pi} 
  \left( \frac{8}{3} f_{3}(t) + \frac{8}{15} f_{1}(t) \right) 
\\
  M^{2}_{\tilde{Q}_{1}}(t) 
&=& 
  M^{2}_{0} + \frac{\alpha_{GUT} M^{2}_{1/2}}{4 \pi}
  \left( \frac{8}{3} f_{3}(t) 
       + \frac{3}{2} f_{2}(t) 
	   + \frac{1}{30} f_{1}(t)
  \right) 
\\
  A_{t}(t) 
&=& 
  \frac{A_{0}}{1 + 6 Y_{t}(0) F(t)}
- M_{1/2} \left( H_{1}(t) 
               - \frac{6 Y_{t}(0) H_{2}(t)}{1 + 6 Y_{t}(0) F(t)}
          \right) 
\\
  A_{u}(t) 
&=& 
  \frac{1}{2} \left( A_{0} + A_{t}(t) - M_{1/2} H_{1}(t) \right) 
\\
  A_{d}(t) 
&=& 
  A_{0} 
- \frac{\alpha_{GUT} M_{1/2}}{4 \pi}
  \left( \frac{16}{3} j_{3}(t) 
       + 3 j_{2}(t) 
	   + \frac{7}{15} j_{1}(t) 
  \right) 
\\
  A_{e}(t) 
&=& 
  A_{0} 
- \frac{\alpha_{GUT} M_{1/2}}{4 \pi}
  \left( 3 j_{2}(t) + \frac{9}{5} j_{1}(t) \right)
\end{eqnarray}
with
\begin{eqnarray}
  t 
&=& 
  \ln(M_{GUT} / Q)^{2} 
\\
  \beta_{i} 
&=& 
  \frac{\alpha_{GUT}}{4 \pi} b_{i} 
\\
  f_{i}(t) 
&=& 
  \frac{1}{\beta_{i}}
  \left(1 - \frac{1}{(1 + \beta_{i} t)^{2}} \right) 
\\
  j_{i}(t) 
&=& 
  \frac{t}{1 + \beta_{i} t} 
\\
  Y_{t}(t) 
&=& 
  \frac{h^{2}_{t}(t)}{(4 \pi)^{2}} 
\\
  E(t) 
&=&
   \left( 1 + \beta_{3} t \right)^{\frac{16}{3 b_{3}}}
   \left( 1 + \beta_{2} t \right)^{\frac{3}{b_{2}}} 
   \left( 1 + \beta_{1} t \right)^{\frac{13}{9 b_{1}}} 
\\
  F(t) 
&=& 
  \int^{t}_{0} E(s) ds 
\\
  H_{1}(t) 
&=& 
  \frac{\alpha_{GUT}}{4 \pi}
  \left( \frac{16}{3} j_{3}(t) 
       + 3 j_{2}(t) 
       + \frac{13}{15} j_{1}(t) 
  \right) 
\\
  H_{2}(t) 
&=& 
  t E(t) - F(t)
\end{eqnarray}
where $b_{1} = 11$, $b_{2} = 1$, and $b_{3} = -3$.
Some of the equations can be found in \cite{RGE}. 

%
\begin{small}
 
\end{small}

%
%
%
\setlength{\unitlength}{1pt}
%
%
\begin{figure} \begin{center} \begin{picture}(468,129)(0,0)
\put(  0,0){\mbox{\epsfig{file=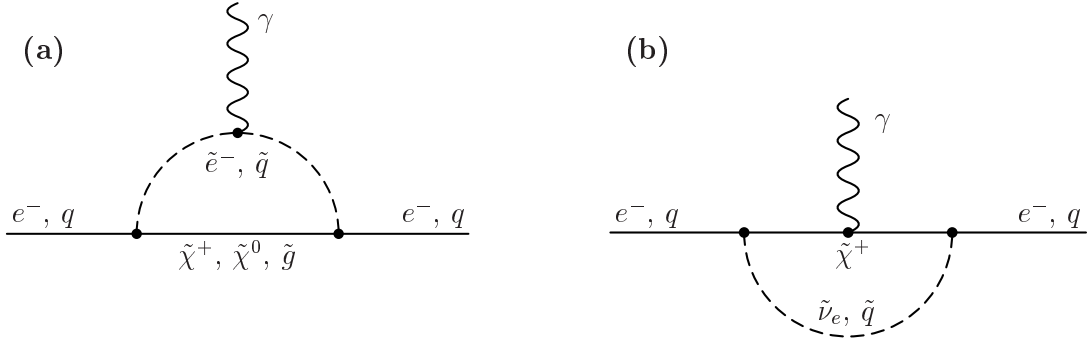,width=468pt}}}
\end{picture}\\ \end{center}
\caption{%
The Feynman diagrams that contribute to the EDMs.
}
\label{fig:feynmangraphs}
\end{figure}
%
%
%
\begin{figure} \begin{center} \begin{picture}(220,165)(0,0)
\put(  0,0){\mbox{\epsfig{file=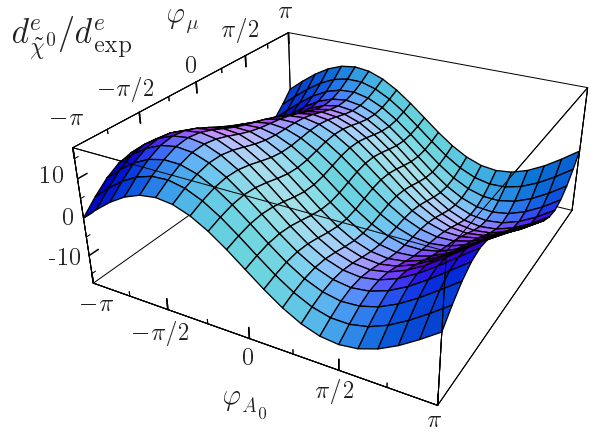,width=220pt}}}
\end{picture}\\ \end{center}
\caption{%
Ratio of the neutralino contribution $d^{e}_{\tilde{\chi}^{0}_{}}$ 
and the experimental limit $d^{e}_{\mathrm{exp}}$ of the electron 
EDM as a function of the phases $\varphi_{\mu}$ and $\varphi_{A_{0}}$. 
The mSUGRA parameters are \mbox{$M_{0} = 150$~GeV}, 
\mbox{$M_{1/2} = 200$~GeV}, 
\mbox{$|A_{0}| = 450$~GeV}, and \mbox{$\tan\beta = 3$}.
}
\label{fig:electron EDM-neutralino}
\end{figure}
%
%
%
\begin{figure} \begin{center} \begin{picture}(220,248)(0,0)
\put(  0,0){\mbox{\epsfig{file=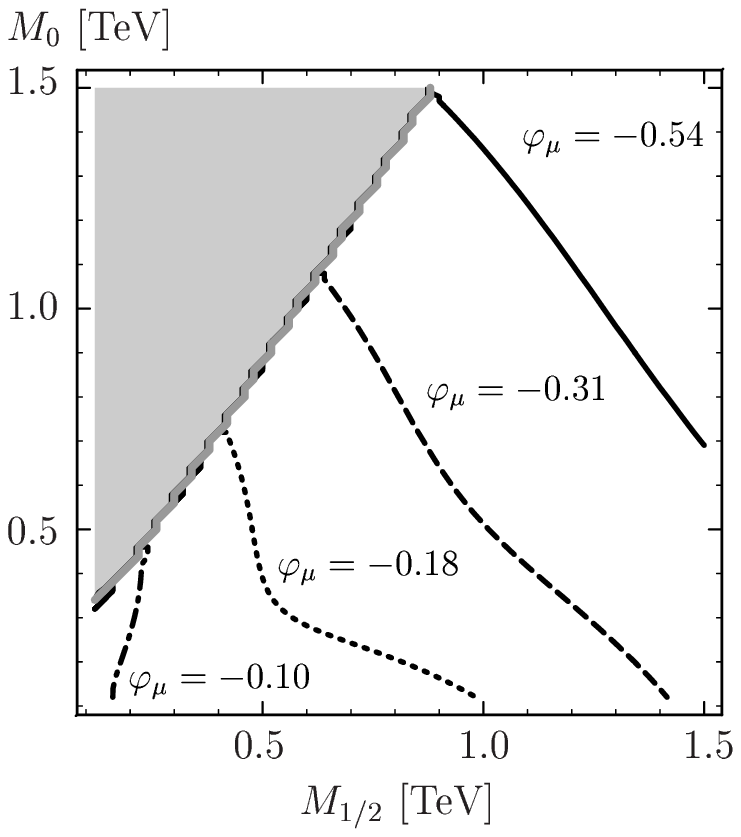,width=220pt}}}
\end{picture}\\ \end{center}
\caption{%
Boundaries of the areas in the $M_{0}$-$M_{1/2}$ plane excluded by 
the electron EDM, for the phases \mbox{$\varphi_{A_{0}} = \pi/2$}, 
and \mbox{$\varphi_{\mu} = -0.54$} (solid line), 
\mbox{$\varphi_{\mu} = -0.31$} (dashed line), 
\mbox{$\varphi_{\mu} = -0.18$} (dotted line), 
\mbox{$\varphi_{\mu} = -0.1$} (dashed--dotted line). 
The areas to the left of the corresponding lines are excluded. 
The mSUGRA parameters are \mbox{$|A_{0}| = 3 M_{0}$} and
\mbox{$\tan\beta = 3$}. 
In the grey area the condition of 
radiative electroweak symmetry breaking is not fulfilled. 
}
\label{fig:electron EDM-mm}
\end{figure}
%
%
%
\begin{figure} \begin{center} \begin{picture}(468,238)(0,0)
\put(  0,0){\mbox{\epsfig{file=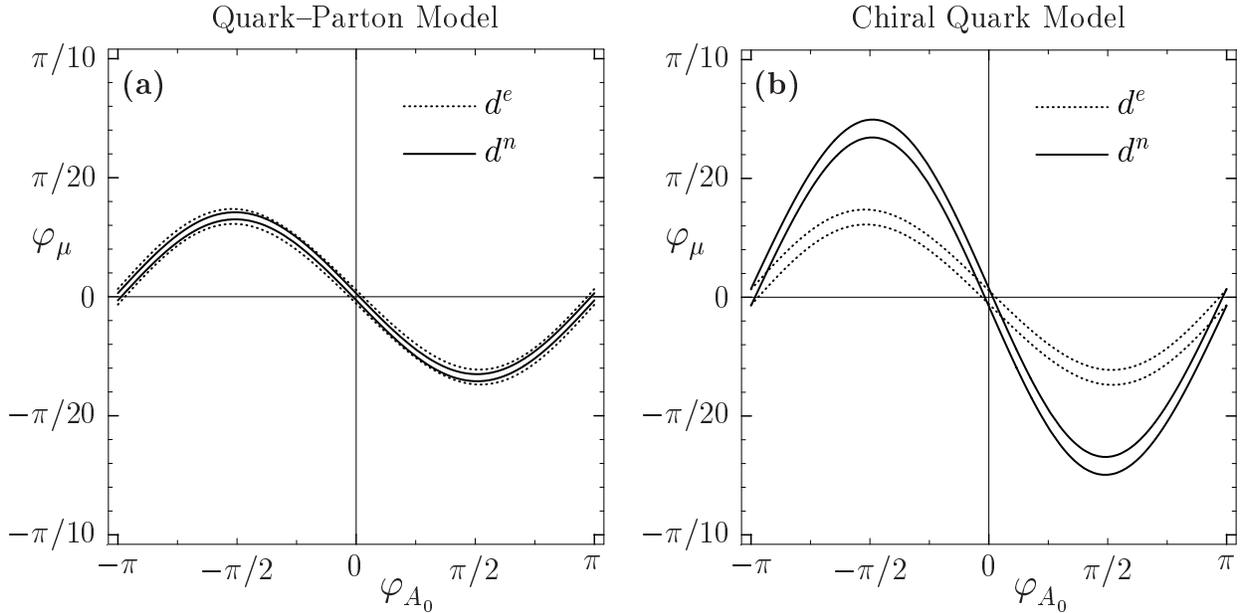,width=468pt}}}
\end{picture}\\ \end{center}
\caption{%
Bands in the $\varphi_{\mu}$-$\varphi_{A_{0}}$ plane allowed by the 
electron EDM (dotted line) and neutron EDM (solid line). 
The mSUGRA parameters are 
\mbox{$M_{0} = 150$~GeV}, \mbox{$M_{1/2} = 200$~GeV}, 
\mbox{$|A_{0}| = 450$~GeV}, and \mbox{$\tan\beta = 3$}. 
The neutron EDM is calculated in the 
Quark--Parton Model \mbox{\textbf{(a)}} 
and in the Chiral Quark Model \mbox{\textbf{(b)}}. 
}
\label{fig:neutron EDM-ph}
\end{figure}
%
%
%
\begin{figure} \begin{center} \begin{picture}(468,248)(0,0)
\put(  0,0){\mbox{\epsfig{file=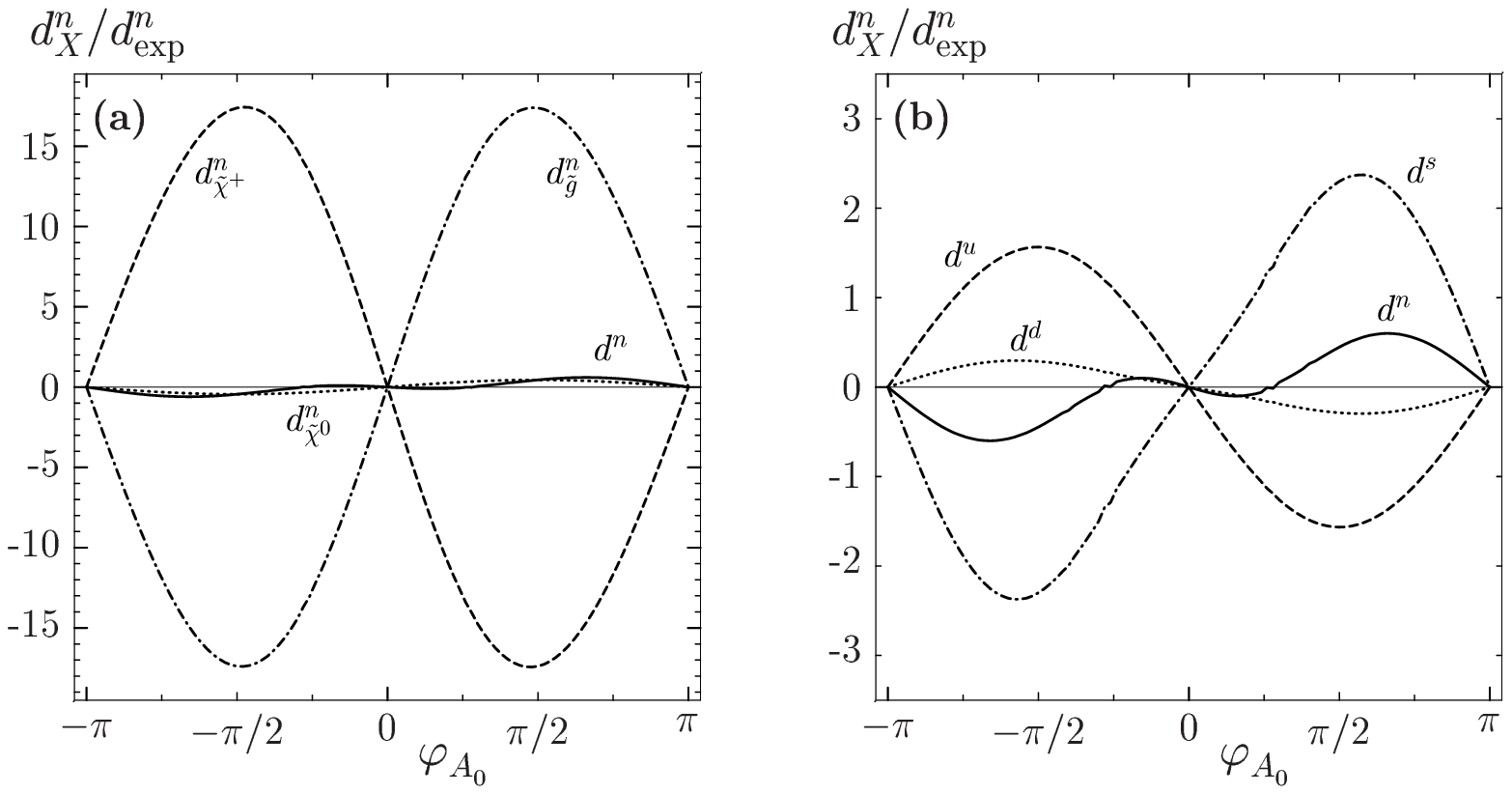,width=468pt}}}
\end{picture}\\ \end{center}
\caption{%
Cancellations of the various contributions to the neutron EDM in the 
Quark--Parton Model, taking the relation 
\mbox{$\varphi_{\mu} = - (\pi/30) \cdot \sin\varphi_{A_{0}}$}.
The mSUGRA parameters are
\mbox{$M_{0} = 150$~GeV}, \mbox{$M_{1/2} = 200$~GeV}, 
\mbox{$|A_{0}| = 450$~GeV}, and \mbox{$\tan\beta = 3$}. 
\mbox{\textbf{(a)}} shows the 
chargino contribution $d^{n}_{\tilde{\chi}^{+}}$ (dashed line), 
neutralino contribution $d^{n}_{\tilde{\chi}^{0}}$ (dotted line), 
gluino contribution $d^{n}_{\tilde{g}}$ (dashed--dotted line), 
and the whole neutron EDM $d^{n}$ (solid line). 
\mbox{\textbf{(b)}} shows the 
up quark contribution $d^{u}$ (dashed line), 
down quark contribution $d^{d}$ (dotted line), 
strange quark contribution $d^{s}$ (dashed--dotted line), 
and the whole neutron EDM $d^{n}$ (solid line). 
}
\label{fig:neutron EDM-cancellation}
\end{figure}
\clearpage
%
%
%
\begin{figure} \begin{center} \begin{picture}(468,248)(0,0)
\put(  0,0){\mbox{\epsfig{file=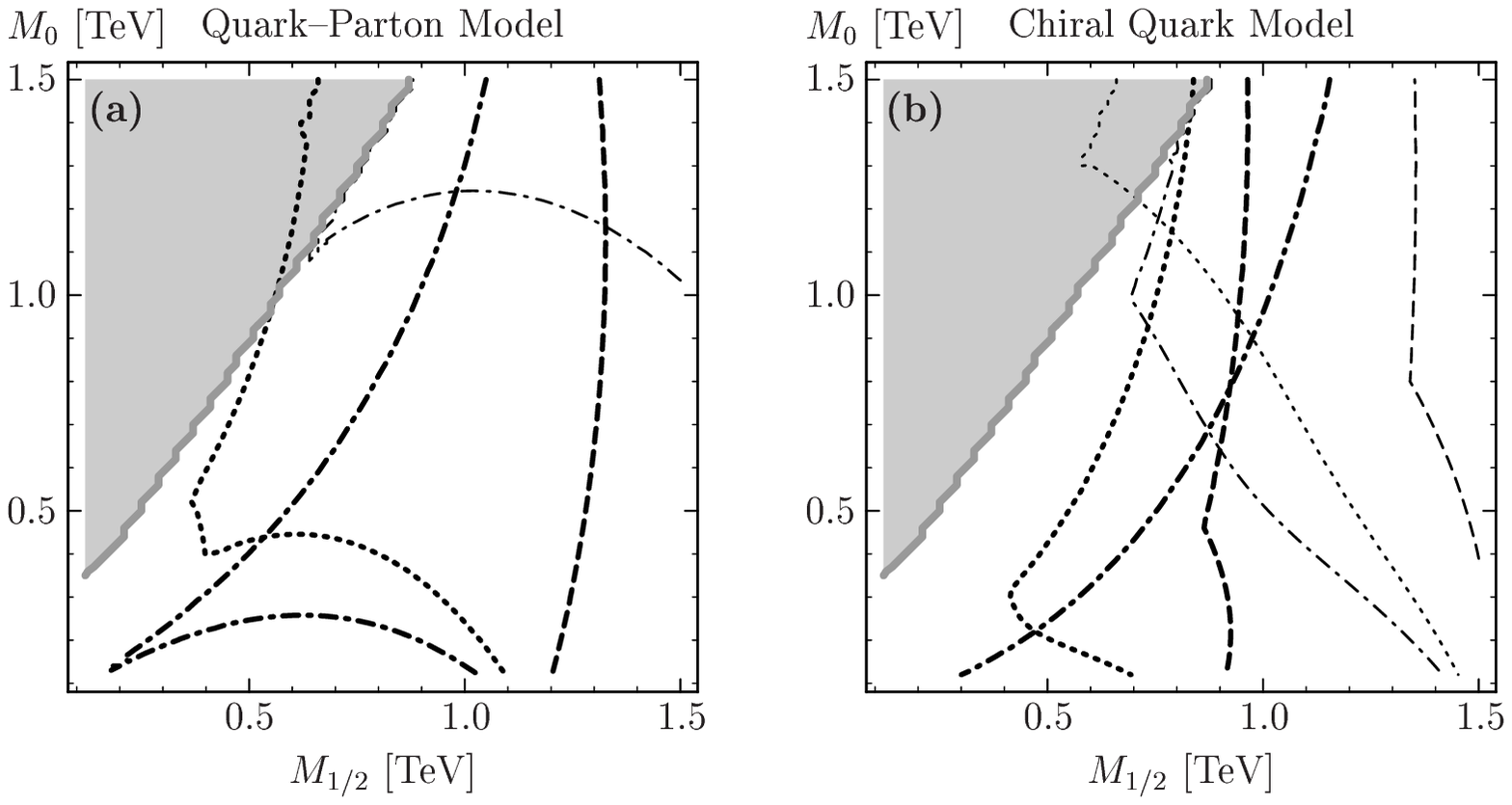,width=468pt}}}
\end{picture}\\ \end{center}
\caption{%
Boundaries of the areas in the $M_{0}$-$M_{1/2}$ plane allowed 
simultanously by the electron EDM and the neutron EDM. The 
neutron EDM is calculated 
in the Quark--Parton Model \mbox{\textbf{(a)}} and
in the Chiral Quark Model \mbox{\textbf{(b)}}. 
The phases are chosen as 
\mbox{$\varphi_{A_{0}} = -\pi/10$} (dashed lines), 
\mbox{$\varphi_{A_{0}} = \pi/5$} (dotted lines),
\mbox{$\varphi_{A_{0}} = \pi/2$} (dashed--dotted lines), and 
\mbox{$\varphi_{\mu} = -\pi/10$} (thin lines), 
\mbox{$\varphi_{\mu} = -\pi/30$} (thick lines). 
The areas to the left and below the corresponding lines are excluded. 
The mSUGRA parameters are \mbox{$|A_{0}| = 3 M_{0}$} and
\mbox{$\tan\beta = 3$}. In the grey area the condition of 
radiative electroweak symmetry breaking is not fulfilled. 
In \mbox{\textbf{(a)}} the whole parameter region is excluded for 
\mbox{$\varphi_{\mu} = -\pi/10$}, \mbox{$\varphi_{A_{0}} = -\pi/10$} 
and 
\mbox{$\varphi_{\mu} = -\pi/10$}, \mbox{$\varphi_{A_{0}} = \pi/5$}.
}
\label{fig:neutron EDM-mm}
\end{figure}
\clearpage
%
%
%
\begin{figure} \begin{center} \begin{picture}(220,230)(0,0)
\put(  0,0){\mbox{\epsfig{file=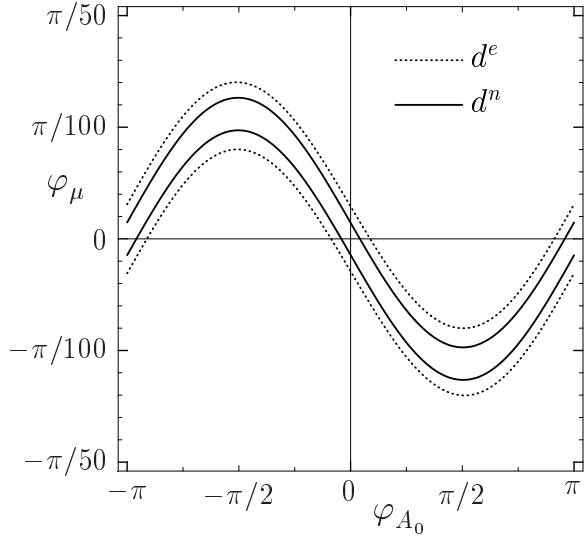,width=220pt}}}
\end{picture}\\ \end{center}
\caption{%
Bands in the $\varphi_{\mu}$-$\varphi_{A_{0}}$ plane allowed by the 
electron EDM (dotted line) and neutron EDM (solid line). 
The mSUGRA parameters are 
\mbox{$M_{0} = 150$~GeV}, \mbox{$M_{1/2} = 200$~GeV}, 
\mbox{$|A_{0}| = M_{0} = 150$~GeV}, and \mbox{$\tan\beta = 3$}. 
The neutron EDM is calculated in the 
Quark--Parton Model. 
}
\label{fig:neutron EDM-smallA0}
\end{figure}
%
%
%
%
\begin{figure} \begin{center} \begin{picture}(220,230)(0,0)
\put(  0,0){\mbox{\epsfig{file=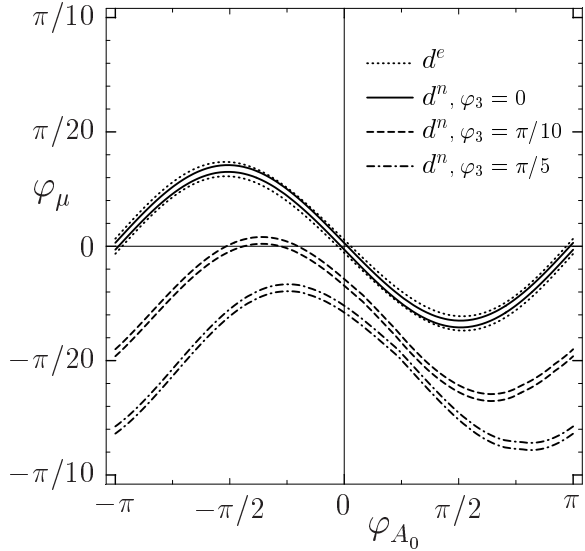,width=220pt}}}
\end{picture}\\ \end{center}
\caption{%
Bands in the $\varphi_{\mu}$-$\varphi_{A_{0}}$ plane allowed by the 
electron EDM (dotted lines) and neutron EDM, calculated 
in the Quark--Parton Model, for \mbox{$\varphi_{3} = 0$} 
(solid lines), 
\mbox{$\varphi_{3} = \pi/10$} (dashed lines), and 
\mbox{$\varphi_{3} = \pi/5$} (dashed--dotted lines).
The mSUGRA parameters are 
\mbox{$M_{0} = 150$~GeV}, \mbox{$M_{1/2} = 200$~GeV}, 
\mbox{$|A_{0}| = 450$~GeV}, and \mbox{$\tan\beta = 3$}. 
}
\label{fig:neutron EDM-ph3}
\end{figure}

%
\begin{table}
\caption{Phases occuring in the mass matrices 
at the electroweak scale and at the GUT scale.}
\begin{tabular}{lll}
mass matrix & electroweak scale & GUT scale \\ 
\hline
$M^2_{\tilde{u}}$ 
 & $\varphi_{\tilde{u}} = \mbox{arg}[A_{u} - \mu^{*} \cot \beta]$
  & $\varphi_{A_{0}}$, $\varphi_{\mu}$ \\
$M^2_{\tilde{d}}$ 
 & $\varphi_{\tilde{d}} = \mbox{arg}[A_{d} - \mu^{*} \tan \beta]$
  & $\varphi_{A_{0}}$, $\varphi_{\mu}$ \\
$M^2_{\tilde{e}}$ 
 & $\varphi_{\tilde{e}} = \mbox{arg}[A_{e} - \mu^{*} \tan \beta]$
  & $\varphi_{A_{0}}$, $\varphi_{\mu}$ \\
$M^{\tilde{\chi}^{+}}$ & $\varphi_{\mu}$ & $\varphi_{\mu}$  \\
$M^{\tilde{\chi}^{0}}$ & $\varphi_{\mu}$ & $\varphi_{\mu}$  \\
\end{tabular}
\label{tab:phases}
\end{table}

\begin{table}
\caption{Values of the phases at the electroweak scale for 
$M_{1/2}$ real and $A_{0} = i A$ imaginary at the GUT scale
when $A = x \, M_{1/2}$.
}
\begin{tabular}{lcccc}
x
& $\varphi_{A_{t}}|_{y = 0.85 (1)}$ 
& $\varphi_{A_{u}}|_{y = 0.85 (1)}$ 
& $\varphi_{A_{d}}$
& $\varphi_{A_{e}}$
\\
\tableline
$0.1$
& $-0.007$ $(0)$
& $-0.021$ $(-0.018)$
& $-0.0278$
& $-0.142$
\\
$1$
& $-0.075$ $(0)$
& $-0.203$ $(-0.177)$
& $-0.271$
& $-0.960$
\\
$10$
& $-0.644$ $(0)$
& $-1.236$ $(-1.190)$
& $-1.207$
& $-1.501$
\end{tabular}
\label{tab:weak-phases}
\end{table}
%
%
\end{document}